\documentclass[a4paper,fleqn,usenatbib]{mnras}

\usepackage{newtxtext,newtxmath}
 
\usepackage[T1]{fontenc}
\usepackage{ae,aecompl}

\usepackage{graphicx}	
\usepackage{amsmath}	
\usepackage{amssymb}	

\usepackage{makecell}
\usepackage{natbib}

\title[Young stars in central cluster galaxies]{Young stellar population gradients in central cluster galaxies from NUV and optical spectroscopy}

\author[Salvador-Rusi\~nol et al.]{
N. Salvador-Rusi\~nol,$^{1,2}$\thanks{E-mail: nsalva@iac.es}
M. A. Beasley,$^{1,2}$
A. Vazdekis $^{1,2}$
and F. La Barbera$^{3}$
\\
$^{1}$Instituto de Astrof\'{i}sica de Canarias, E-38200 La Laguna, Tenerife, Spain\\
$^{2}$Departmento de Astrof\'{i}sica, Universidad de La Laguna, E-38205 La Laguna, Tenerife, Spain\\
$^{3}$INAF – Osservatorio Astronomico di Capodimonte, I-80131 Napoli, Italy
}

\date{Accepted XXX. Received YYY; in original form ZZZ}

\pubyear{2020}

\begin{document}
\label{firstpage}
\pagerange{\pageref{firstpage}--\pageref{lastpage}}
\maketitle

\defcitealias{labarbera2019}{LB19}
\defcitealias{SR19}{SR19}

\begin{abstract}

Central cluster galaxies are the largest and most massive galaxies in the Universe. Although they host very old stellar populations, several studies found the existence of blue cores in some BCGs indicating ongoing star formation. We analyse VLT/X-Shooter stacked spectra of 6 nearby massive central galaxies with high central velocity dispersions ($\sigma$>300 kms$^{-1}$) at different galactocentric distances. We quantify the young stellar population out to 4 kpc by fitting near-UV and optical absorption line indices with predictions of composite stellar populations modelled by an old and a young stellar component. We also use IMF-sensitive indices since these galaxies have been found to host a bottom-heavy IMF in their central regions. We derive negative young stellar populations gradients, with mass fractions of stars younger than 1 Gyr decreasing with galactocentric distance, from 0.70\% within 0.8 kpc to zero beyond 2 kpc. We also measure the mass fraction in young stars for individual galaxies in the highest S/N central regions. All the galaxies have young components of less than one percent. Our results clearly suggest that the star formation in massive central cluster galaxies takes place in their galaxy cores (<2 kpc), which, with deeper gravitational potential wells, are capable of retaining more gas. Among the possible sources for the gas required to form these young stars, our results are consistent with an in-situ origin via stellar evolution, which is sufficient to produce the observed young stellar populations.
\end{abstract}




\section{Introduction}


Most galaxy clusters are dominated by a remarkable class of elliptical galaxies commonly located at the cluster core (\citealt{ostriker1975}, \citealt{hausman1978},  \citealt{postman1995}). They are usually the brightest and the largest galaxy in a cluster, and are commonly referred as Brightest Cluster Galaxies (BCGs). BCGs are the most massive and luminous galaxies in the present-day Universe, differing from normal ellipticals for their extended brightness distributions.

Like normal elliptical galaxies, the analysis of the stellar populations in the inner regions of BCGs indicates that the bulk of their stars was formed rapidly in a very intense star burst at redshift $z>2$ \citep{renzini2006}. However, despite sharing similar morphologies with normal massive ellipticals, in addition to red colours, old and metal-rich stellar populations and alpha-enhancements (\citealt{brough2008}, \citealt{loubser2009}, \citealt{donahue2010}, \citealt{loubser2012}, \citealt{barbosa2016}, \citealt{edwards2020}), BCGs constitute a special category of objects with peculiar star formation histories (SFHs) seen from both observations (\citealt{tran2008}, \citealt{barbosa2016}) and models (\citealt{dubinski1998}, \citealt{delucia2007}). The evolution of BCGs is significantly affected by their surrounding environments. Due to their central positions in the gravitational potential well of their host clusters, central cluster galaxies accrete stars and gas from satellite galaxies that orbit around them and fall in, developing extended light profiles. The more representative SFHs of their dominant stellar population include components from in-situ star formation, and from the interaction with other galaxies and with the intra-cluster medium, where the outer regions in BCGs are continually assembling mass through minor mergers \citep{cooke2019}. The stellar populations of a large sample of observed BCGs have been recently studied in \cite{edwards2020}, from the galaxy core into the intra-cluster light (ICL) out to 4 effective radii, finding old stellar populations of $\sim13$ Gyr and high metallicities [Fe/H]~$\sim0.3$ in the galaxy cores, whereas the average age in the ICL is estimated to be slightly younger, $\sim9.2$ Gyr with lower metallicities --0.4 < [Fe/H] < 0.2 at 40 kpc. This broadly supports the idea of two-phase galaxy formation, with the BCG cores and inner regions formed faster and earlier than the outer regions that have  formed more recently, or have accreted mass afterwards through galaxy mergers and thereby also  increasing  in size (\citealt{oser2010}, \citealt{kubo2017}, \citealt{cooke2019}). 

Based on their old ages and high [Mg/Fe] abundances, the star formation in these massive systems needs to be rapidly quenched early, at high redshift. The removal of gas reservoirs required to form new stars could be due to different processes: ram pressure stripping by the intracluster medium, tidal disruption with the ICL, galaxy mergers and interactions with other galaxies, feedback from active galactic nucleus (AGN) from supermassive black holes \citep{binney1995}. Even though all these mechanisms might contribute to some extent to physically remove the gas reservoirs, there is a general agreement for AGN feedback to be a key agent at suppressing the star formation in massive galaxies (\citealt{schawinski2007}, \citealt{weinberger2017}, \citealt{scholtz2018}).


The star formation in BCGs in the local Universe is much lower than what used to be at higher redshifts (z>2) where the bulk of the stars was formed. Several studies provide important insights of recent star formation in some low redshift BCGs (\citealt{crawford1993}, \citealt{cardiel1998}, \citealt{crawford1999}, \citealt{edge2001}, \citealt{odea2008}, \citealt{bildfell2008}, \citealt{pipino2009}, \citealt{odea2010}, \citealt{hicks2010}, \citealt{hoffer2012}, \citealt{liu2012}, \citealt{fraser-mcKelvie2014}, \citealt{cooke2016}, \citealt{loubser2016}, \citealt{runge2018}, \citealt{cerulo2019}). In particular, using GALEX \citep{martin2005} UV data, \cite{pipino2009} found the existence of blue cores in a sample of 7 BCGs as evidence of recent star formation. Specifically, their young component is younger than 200 Myr and contribute less than one percent to the total stellar mass. \cite{liu2012} also find similar ongoing star formation in BCGs, whose level of star formation is higher for more massive BCGs located in richer galaxy clusters.

The mechanisms that trigger star formation in BCGs have been a matter of study in the last two decades. In central cluster galaxies with high velocity dispersions, the origin of the cold gas necessary to form new stars is unlikely to originate from wet (i.e. gas-rich) mergers in the present Universe. Such mergers are rare between galaxies with high relative velocity dispersion, such as BCGs and their surrounding galaxies, and both observations and semi-analytic approaches seem indicate that they were common at high redshift, whearas at low redshift dry mergers (i.e. gas-poor) dominate (\citealt{lidman2012}, \citealt{edwards2012}, \citealt{hilz2013}, \citealt{lavoie2016}, \citealt{stockmann2020}). Several studies have shown that BCGs that experience recent star formation are commonly located in clusters with the presence of cooling flows from the ICM (\citealt{bildfell2008}, \citealt{rafferty2008}, \citealt{donahue2010}, \citealt{liu2012}, \citealt{loubser2016}, \citealt{cerulo2019}). \cite{bildfell2008} investigated BCG colours in a wide variety of clusters X-ray morphologies, including cool-core and non-cool core clusters. They show that 25\% of their BCG sample exhibit colour profiles that become bluer towards the galaxy core and they associate this with ongoing star formation. They suggest that this recent star formation is linked to environmental process within the cluster, as they find that these blue-core BCGs are all located within $\sim$10 kpc distance to the cluster X-ray peak, where the cooling flow is preferentially going to and the cold gas of the ICM is preferentially deposited. More recently, in agreement with \cite{bildfell2008}, \cite{cerulo2019} show that this cool-core process that fuel the star formation in BCGs is not related to the halo mass of the clusters.


In this context, the present work is motivated by a recent study which has shown that, by measuring near-ultraviolet (NUV) line-strength indices from a large sample of SDSS ETG spectra, massive ETGs at z$\sim$0.4 are still forming new stars at very low levels (\citealt{SR19}, hereafter \citetalias{SR19}). \citetalias{SR19} found that massive ETGs have mass fractions of $\sim$0.5\% of stars younger than 2 Gyr within their 10 kpc central regions. In this paper we present an analysis of young stellar population gradients in massive central galaxies from spectra observed at different galactocentric distances. Our analysis is based on high-quality spectroscopy taken with the X-Shooter spectrograph at VLT in the UV and optical spectral ranges, for a sample of six massive BCGs (with velocity dispersions $\sigma$>300 kms$^{-1}$) at redshift z$\sim$0.05. We take advantage of the sensitivity of NUV absorption lines to young stars. NUV indices, relative to the optical, are extremely sensitive to the youngest stellar component of stellar systems like BCGs, since they allow also tiny fractions (<1\%) of young stars to be detected in galaxy integrated spectra (\citealt{vazdekis2016}, \citetalias{SR19}). We compare observed absorption line-strength indices to predictions from the E-MILES models. Hence, we aim to quantify the radial content of young stars in the central regions of massive central galaxies to up to $\sim$ 4 kpc distance from the galaxy centre. Intimately connected to their environment, BCGs are very special objects to study how their young stars are spatially distributed. Due to their high gravitational potential wells, they are expected to have a larger concentration of gas towards the galaxy central regions. Thus, their cores are expected to have greater star formation activity than in their outer regions.

The paper is organized as follows: in Section \ref{sec:data}, we present the data used and the stacking process. Section \ref{sec:methods} describes the set of absorption line-strength indices and the stellar population models used to fit them. Here we also describe our modelling approach to infer the young stellar component. We derive the young stellar gradients and the stellar population parameters in Section \ref{sec:results}. Finally, the discussion and conclusions are summarized in Section \ref{sec:discussion}. Throughout this paper, we assume a Lambda-Cold Dark Matter Universe with values of Hubble constant H$_{0}$ = 70 kms$^{-1}$Mpc$^{-1}$, matter density parameter $\Omega_{m}$ = 0.3 and a cosmological constant $\Omega_{\Lambda}$ = 0.7.


\section{Data} \label{sec:data}

\subsection{Galaxy sample}

\begin{figure*}
	\includegraphics[width=\textwidth]{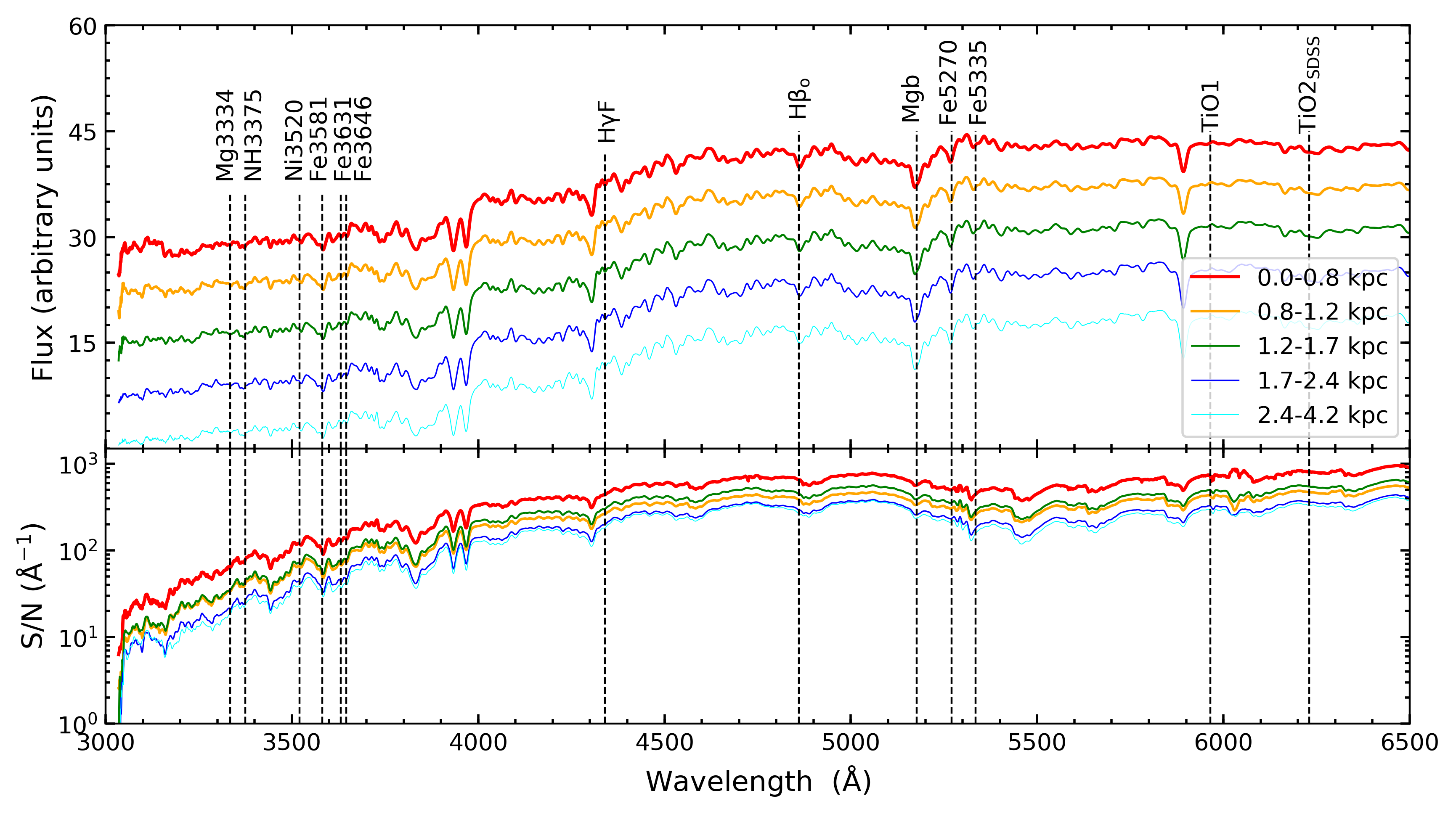}
    \caption{The stacked spectra of our five radial bins. In the upper panel we show the stacked spectra in the range of interest of this work, 3000 -- 6500 \AA. The lower panel shows the S/N per \AA \ of each spectrum. Both panels are color-coded according to their radial bin. Spectra are smoothed at a common velocity dispersion of 350 kms$^{-1}$. Dashed lines indicate the central wavelength of the spectral indices used in the analysis.}
    \label{fig:spectra}
\end{figure*}

The data used in this paper consist of galaxy spectra from the VLT/X-Shooter spectrograph \citep{vernet2011} of six very massive central cluster galaxies at different galactocentric distances, covering from 0 to $\sim$4.2 kpc, within redshift 0.048 $\leq$ z $\leq$ 0.056. We use the same dataset used in \cite{labarbera2019} (hereafter \citetalias{labarbera2019}), except for one target, named XSG2 (see below), that has been classified as a satellite galaxy, and therefore it is not included in our sample of BCGs. Notice that the main scientific goal was to explore the radial gradients of the stellar initial mass function (IMF) in a set of very massive ETGs. We refer the reader to \citetalias{labarbera2019} for a detailed description of how the observations and data reduction of these data have been performed. Briefly, the X-Shooter spectra consist of data taken with three independent arms, ultraviolet-blue (UVB), visible (VIS) and and near-infrared (NIR). The data were obtained with 0.9'', 0.9'', and 1.0'' slit widths in UVB, VIS, and NIR, respectively. The wavelength range of the final, joined, spectra is $\sim$ 3000 -- 23500 \AA. In the present work, we focus our analysis from the bluest (NUV) to optical wavelengths. For our purposes, we refer to NUV as the spectral range between 3000 -- 3700 \AA. Observations were carried out in service mode and have typical exposure times on each target of 1.7, 1.9, and 2.1 hours, in the UVB, VIS, and NIR arms, respectively, with a median seeing of $\sim$0.8-0.9 arcsec. The spectra were extracted at different galactocentric distances. The innermost spectra of all galaxies were extracted within an aperture of width 1.3'' around the photometric centre. The aperture size was then increased outwards, in order to ensure a median S/N > 90 per \AA \ in the optical range of the individual galaxy spectra. These spectra have been corrected for Galactic reddening and shifted to rest-frame accordingly.

Following the same nomenclature as \citetalias{labarbera2019}, we refer to our galaxies as XSG1, XSG6, XSG7, XSG8, XSG9 and XSG10\footnote{Galaxy names according to SDSS are J142940.63+002159, J144120.36+104749.8, J151451.68+101530.4, J015418.07-094248.4, J005551.88-095908.3 and J075354.98+130916.5, respectively. General galaxy properties are summarised in Table 1 of \citetalias{labarbera2019}.}. Our sample excludes the XSG2 galaxy that is the only genuine satellite according to the analysis presented in \citetalias{labarbera2019}, but we analyse its young stellar contribution in Appendix \ref{app:xsg2}. Five galaxies were selected from the lowest redshift limit (z $\sim$ 0.05) of the SPIDER sample \citep{labarbera2010a}, while one target (XSG10) was selected from SDSS-DR7 applying the same criteria as SPIDER ETGs, but at slightly lower redshift (z$\sim$ 0.048). This sample was selected to comprise very massive galaxies, so they have very high central velocity dispersions (>300 kms$^{-1}$ for the innermost spectra for all galaxies). With the exception of XSG7, all targets have been classified as the central galaxy of a group or cluster of galaxies, i.e. BCGs. However, relevant for the present work, is that this galaxy has similar stellar masses to those of the brightest galaxy of their parent groups. Also, XSG7 is also located at the central region of a group with another bright companion classified as group central galaxy. Therefore, although not being classified as a central cluster galaxy, it was likely central to the  infallen group for most of its evolution (see \citetalias{labarbera2019} for more details). BCGs are generally located at the cluster centroid, but there do exist some cases of BCGs located away from the centre of their host cluster \citep{lidman2013}. The main result from \citetalias{labarbera2019} is that these galaxies show negative IMF gradients, with bottom heavy IMFs ($\Gamma_{b}\sim$2.8--3.5) in the innermost galaxy regions which changes to a standard logarithmic slope $\Gamma_{b}\sim$1.3 at galactocentric distances of $\sim$4 kpc. Also, all these galaxies have high [Mg/Fe] abundances, from $\sim$0.3 to 0.5 dex (\citetalias{labarbera2019}), indicating short star formation timescales early in the Universe. Hence, the galaxy sample studied here are low-redshift, very massive galaxies corresponding to the extreme high-mass end of ETGs, located at the centres of galaxy clusters.


\subsection{Stacking the spectra}

Massive galaxies, such as those in our sample, are known to be dominated by very old stellar populations. Therefore, detecting the contribution of the youngest stars in their spectra is difficult. The NUV luminosity of galaxies is a good tracer of young stellar populations, but requires high S/N spectra. Unfortunately, obtaining high S/N spectroscopy in the NUV is not an easy task, mostly due to the faint flux level of early-type galaxies in this spectral range, and also because of the significant absorption from the atmosphere. Despite the very high-quality of the data presented here, the individual galaxy spectra for our sample still have relatively low S/N in the NUV range. The S/N is, on average, 13 \AA$^{-1}$ for the innermost galaxy spectra, which is sufficient for detecting the presence of small amounts of young stars from the NUV spectral indices. However,  at the outermost galactocentric distances it is much lower ($\sim$ 3 \AA$^{-1}$).  Therefore, we take advantage of the fact that all the galaxies in our sample are massive, with very similar central velocity dispersions, to increase the S/N by stacking the spectra into a single spectrum according to their physical radial distance to the galaxy centre within 4.2 kpc. In order to explore the behaviour of the young stellar component with distance from the galaxy centre, the individual spectra are grouped into five bins according to their galactocentric distances. These bins cover 0 -- 0.8 kpc, 0.8 -- 1.2 kpc, 1.2 -- 1.7 kpc, 1.7 -- 2.4 kpc and 2.4 -- 4.2 kpc radial ranges. The individual galaxy spectra are stacked together in each bin, so that all the galaxy sample contribute to each radial bin. Note that the outermost radial bin has slightly larger size in order to ensure a S/N>13 per \AA \ in the NUV range in the stacked spectra, necessary to measure certain spectral indices. We create the stacks by bringing each individual galaxy spectrum to rest-frame wavelength and taking the median flux at each common wavelength as the stacked flux value. The error on the flux comes from the standard deviation of the individual galaxy fluxes at each wavelength. This way, we have five spectra of high S/N at different radial distances representative of very massive central galaxies. 

The final stacked spectra with their corresponding S/N are shown in Fig. \ref{fig:spectra}. The central wavelength of the spectral indices used in our analysis are identified by vertical dashed lines. The stacked spectra have high S/N, above $\sim$15 at wavelengths larger than 3300 \AA \ and several hundreds in the optical. In particular, the central bins have higher S/N than the outermost ones. This permits us to increase the number of absorption spectral lines usable for the analysis of the innermost bins. All spectra exhibit relatively modest NUV fluxes compared to the optical range and no strong emission lines are clearly seen, indicative of the presence, on average, of quiescent stellar populations in our galaxies. Although some of our targets show some emission contamination in their individual spectra (see App. F in \citetalias{labarbera2019} for a detailed analysis) the final stacked spectra are not corrected for the emission, since our index sample of interest is not affected by the emission contamination.

Note that the index values depend on the resolution of the spectra and on the velocity dispersion of the galaxy at a given bin. Therefore, when comparing with the theoretical predictions, both models and observations need to be at the same spectral resolution. Before the stacking, we performed gaussian smoothing in both models and individual galaxy spectra at a common velocity dispersion of 350 km$^{-1}$. This is the highest velocity dispersion of all galaxy spectra, which were measured with the software pPXF \citep{cappellari2004}.


\section{Methodology} \label{sec:methods}

In this section we outline the methodology used to estimate the young stellar component for these massive galaxies. Our method consists of fitting the observed spectral lines with stellar population synthesis model predictions that are derived from a simple modelling of the SFH for very massive galaxies. This method is based on that developed in \citetalias{SR19}.

\begin{table*}
 \centering
 \begin{tabular}{|c|c|c|c|c|}
 \hline	
 Index & Blue Passband (\AA) & Index Passband (\AA) & Red Passband (\AA) & Reference\\
  (1) & (2) & (3) & (4) & (5) \\
 \hline
  Mg3334 & 3310.000 -- 3320.000& 3328.000 -- 3340.000 & 3342.000 -- 3355.000 & \cite{serven2011} \\
 NH3375 & 3342.000 -- 3352.000& 3350.000 -- 3400.000& 3415.000 -- 3435.000 & \cite{serven2011} \\
 Ni3520 & 3499.000 -- 3508.000& 3511.000 -- 3530.000& 3532.000 -- 3548.000 & \cite{gregg1994} \\
  Fe3631 & 3625.500 -- 3628.500 & 3628.500 -- 3637.500& 3659.000 -- 3674.000 & \cite{gregg1994} \\
  Fe3646 & 3625.500 -- 3628.500 & 3642.500 -- 3659.000& 3659.000 -- 3674.000 & \cite{gregg1994} \\  
 H$\gamma$F & 4283.500 -- 4319.750 & 4331.250 -- 4352.250 & 4354.750 -- 4384.750 & \cite{worthey1997} \\
 H$\beta_{o}$ & 4821.175 -- 4838.404 & 4839.275 -- 4877.097 & 4897.445 -- 4915.845 & \cite{cervantes2009} \\
 Mgb5177 & 5142.625 -- 5161.375 & 5160.125 -- 5192.625 & 5191.375 -- 5206.375 & \cite{worthey1994} \\
 Fe5270 & 5233.150 -- 5248.150 & 5245.650 -- 5285.650 & 5285.650 -- 5318.150 & \cite{worthey1994} \\
 Fe5335 & 5304.625 -- 5315.875 & 5312.125 -- 5352.125 & 5353.375 -- 5363.375 & \cite{worthey1994} \\
TiO1 & 5816.625 -- 5849.125 & 5936.625 -- 5994.125  & 6038.625 -- 6103.625 & \cite{worthey1994} \\
TiO2$\rm_{SDSS}$ & 6066.625 -- 6141.625 & 6189.625 -- 6272.125 & 6422.000 -- 6455.000 & \cite{labarbera2013} \\
 \hline
 \end{tabular}
\caption{Definitions of the NUV and optical spectral indices used in this study. Each index listed in the column 1 is defined by a central passband (column 3) and a blue and a red pseudo-continuum spectral range (columns 2 and 4, respectively). References of the definitions are listed in the column 5. The combined optical index [MgFe]$^\prime$ is obtained via:
 [MgFe]$\rm^\prime=\sqrt{Mgb5177(0.72Fe5270+0.28Fe5335)}$ \citep{thomas2003}.
   }
 \label{tab:indices}
\end{table*}

\subsection{Line-strength indices in the NUV and optical}

Thanks to the wide spectral range of the X-Shooter spectrograph, we are able to use spectral indices from NUV to optical wavelengths. This allows us to constrain the various stellar components that contribute with different proportions to these spectral ranges. Definitions of the indices used in this study are listed in Table \ref{tab:indices}. Line-strength index fitting has been used by a large number of spectroscopic studies to obtain stellar population properties of massive galaxies (e.g., \citealt{sanchezblazquez2006b}, \citealt{peletier2007}, \citealt{brough2007}, \citealt{loubser2009}, \citealt{zhu2010}, \citealt{harrison2010}, \citealt{choi2014}, \citealt{ferremateu2014}, \citealt{francois2019}). In this study, we use features in the NUV range that are sensitive to very small mass fractions of young stars (\citealt{vazdekis2016}, \citetalias{SR19}) that allow us to constrain the youngest stellar populations of massive old galaxies. We note that wavelengths larger than 3000 \AA \ are less sensitive to young stars in comparison to bluer wavelengths. However, with sufficient S/N, this range can also be a good tracer of the youngest stellar populations. Indices in the NUV spectral range are selected in order to have large sensitivity to small fractions of young stars  (< 1 Gyr). Particularly, we choose NUV indices that are most sensitive to these components in comparison to other effects such as alpha-element enhancement, which is observed in these galaxies. For this purpose, we have explored how each index in the NUV spectral range is influenced by the presence of a small mass fraction (0.1\%; see \citetalias{SR19}) of a young stellar population on top of an old component, and by an alpha-enhancement of [Mg/Fe]=0.4. For the latter we employ the \cite{vazdekis2015} and \cite{conroy2012} stellar population models, which use the MILES stellar library \citep{sanchezblazquez2006} and differential corrections from theoretical star spectra. We note that the NH3375, Mg3334 and Ni3520 indices are outside the wavelength range where these models are defined and provide safe predictions. Thus, we use them since they are the bluest indices in the spectral range of our spectra and are therefore the most sensitive to the youngest stars. We have excluded Fe3581, Fe3619, Fe3683 and Fe3706 indices since the models suggest they are strongly affected by the Mg/Fe abundance, which leads to variations of similar magnitude to those from the young component. The Fe3741 absorption line is excluded for being affected by the oxygen emission line in 3727 \AA, whose impact, despite being small, it is not completely negligible. We include the Fe3631 and Fe3646 indices which, again according to the models, appear to be more sensitive to the presence of young stars than [Mg/Fe] abundance.

We also include indices within the optical range that trace the bulk of the stellar population. We use age-sensitive Balmer lines and particularly, H$_{\beta_{o}}$, which  provides better constraints on the age than the classical Lick index H$_{\beta}$ \citep{cervantes2009}. We include [MgFe]', which is a combination of Mg and Fe indices, defined to be insensitive to variations of [Mg/Fe] abundance ratio \citep{thomas2003} but sensitive to the total metallicity. We also include the TiO1 and TiO2$\rm_{SDSS}$ indices for being strongly sensitive to the IMF slope, as our galaxies have been shown to have a strong IMF gradient. These optical indices have been widely used in a large number of stellar population studies of massive galaxies. 

Since the S/N in the NUV range decreases with increasing radius, some indices are excluded in the outer radial bins in our analysis because of their S/N requirements. Mg3334 can only be used in the two central bins, since it requires S/N>40 per \AA \ for being able to disentangle small fractions (<1\%) of young stars on top of old stellar populations. Similarly, the Fe3631 and Fe3646 indices are not used in the outermost bin. Therefore, the set of spectral indices used in our analysis changes with each galactocentric distance bin as summarized in Table \ref{tab:indexset}. However, to make sure that our results are robust, we have checked that the omission of these indices does not introduce any systematic effects in our results. This has been assessed by deriving the young stellar component by including (and excluding) these indices during the fitting process in all radial bins, and finding that the best-fit solution does not change and thus, neither does the observed trend with galactocentric distance.

The error associated to the index measurements comes from two different sources. First, from the Poissonian noise, which is smaller for the optical indices as a result of the higher S/N of the stacked spectra in this wavelength range. Second, the error on the stacked spectra is measured by stacking the errors on the fluxes of the individual galaxy spectra. This error is included in order to account for the variations among the six galaxies in a given stack, since the poissonian noise is very small and does not represent the individual galaxy indices. Finally, the adopted index uncertainties come from the Poissonian noise added in quadrature to the noise from the individual galaxy flux errors.

\begin{table}
 \centering
 \begin{tabular}{ccc}
 \hline	
Bin (kpc) & $\Gamma_{b}$ & Set of indices used \\
  (1) & (2) & (3) \\
 \hline
0.0 -- 0.8  & 3.0 & \makecell{Mg3334, NH3375, Ni3520, Fe3631, \\ Fe3646, H$\gamma$F, [MgFe]$^\prime$, TiO1, TiO2$\rm_{SDSS}$} \\
0.8 -- 1.2 & 2.8 & Same as first bin \\
1.2 -- 1.7 & 2.5 & Same as first bin but w/o Mg3334 \\
1.7 -- 2.4 & 1.8 & Same as first bin but w/o Mg3334 \\
2.4 -- 4.2 & 1.3 & Same as first bin but w/o Mg3334, Fe3631 and Fe3646 \\
 \hline
 \end{tabular}
\caption{Properties of the stacked spectra. Column 1 indicates the galactocentric distance range of each bin. Column 2 lists the average logarithmic IMF slope of each radial bin from \citetalias{labarbera2019}. Column 3 summarizes the index sample used in the fitting process in each bin for constraining the young stellar component.
   }
 \label{tab:indexset}
\end{table}

\subsection{Stellar population models} \label{sec:sspmodels}

To analyse both the NUV and optical indices we make use of the E-MILES single stellar population (SSP, i.e., a population of stars born at the same time with the same metallicity) models described in \cite{vazdekis2016}. These models are based  on empirical stellar libraries covering from the UV (NGSL, \citealt{gregg2006}), optical (MILES, \citealt{sanchezblazquez2006}) to the near-infrared (Indo-US, \citealt{valdes2004}, CAT, \citealt{cenarro2001a}, IRTF, \citealt{cushing2005}, \citealt{rayner2009}) wavelength ranges. The models are computed for different IMF shapes and slopes. In particular, we adopt a low-mass tapered "bimodal" IMF \citep{vazdekis1996} with varying logarithmic slopes. These models span an age grid from 63 Myr to 14 Gyr, and select a metallicity grid that ranges from [M/H] = $-0.40$ to 0.22 dex. Guided by the results obtained in LB19, we have performed a linear extrapolation of the metallicity out to [M/H]=0.45 to be able to constrain properly the amount of young stellar component in the innermost galactocentric radial bin. For younger ages, we used the version of E-MILES models extended to 6.3 Myr \citep{asad2017}. We therefore use the version constructed with Padova isochrones \citep{girardi2000}, since the youngest SSP models are only available for this set of isochrones. Line-strength index predictions from the theoretical models as given by the approach defined in Section \ref{sec:sfhparam} are compared to those measured in the observed stacked spectra. We estimate the young stellar component at each radial bin by varying age, metallicity and IMF. We degraded all the SSP models to the resolution of 350 kms$^{-1}$ to match the common resolution for all the galaxies and galaxy bins. 

\subsection{Constraints on the age and metallicity} \label{sec:mlwa}

\begin{figure}
	\includegraphics[width=\columnwidth]{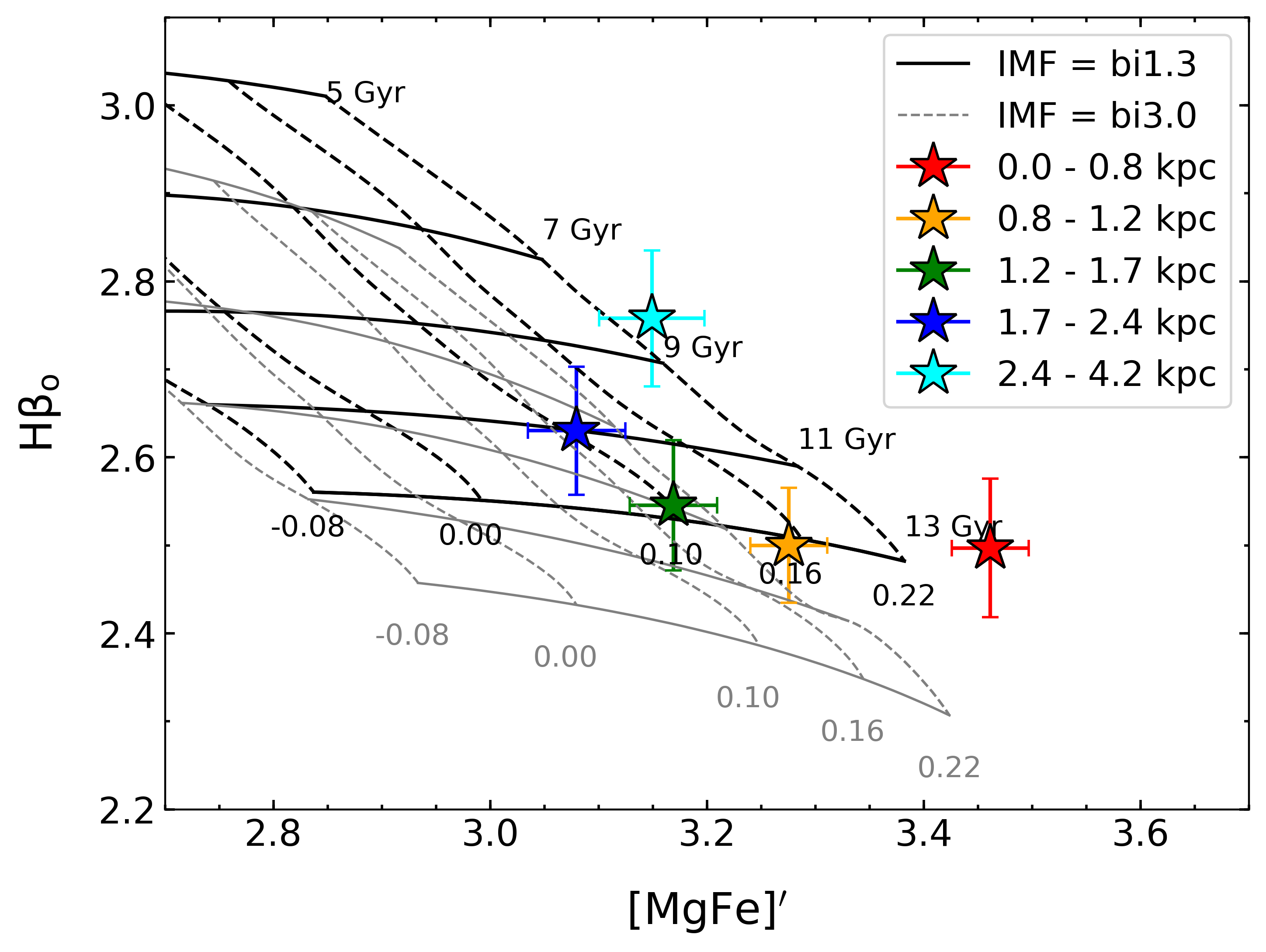}
    \caption{Age-sensitive H$_{\beta_{o}}$ index is plotted versus the total metallicity-sensitive indicator [MgFe]'. Grids show the E-MILES model predictions for a standard IMF (black) and a bottom-heavy IMF (grey) for different ages (solid lines), increasing from top to bottom, and metallicities (dashed lines), which increase from left to right. Stars show values measured in each stacked spectrum, following the same color-code used in Fig. \ref{fig:spectra}. Data and models have been smoothed to the same spectral resolution of $\sigma=$350 kms$^{-1}$.}
    \label{fig:grid}
\end{figure}

An integrated galaxy spectrum from predominantly old stellar systems such as BCGs can be approximated with one SSP in the optical range. We initially derive the mean luminosity-weighted ages and metallicities of the stellar population in the stacked galaxy spectra using two optical indices. Fig. \ref{fig:grid} shows the widely used index-index plot in the visible with the age-sensitive index H$_{\beta_{o}}$ versus the total metallicity indicator [MgFe]'. SSP model predictions are shown for different ages and metallicities for two grids with two different bimodal IMFs. We show the predictions corresponding to a logarithmic standard slope 1.3 in black, which is representative of the outer bins, and for a bottom-heavy IMF with slope 3.0 in grey representative of the inner bins, as derived in \citetalias{labarbera2019}. We overplot the measurements of the stacked spectra (star symbols) that indicate old ages and metal-rich ([M/H]>0.0) stellar populations. The inner regions have higher [MgFe]' values, revealing more metal-rich stellar populations towards the galaxy centre. We also see that the two innermost bins fall outside the standard IMF slope grid, which indicates a  bottom-heavy (dwarf-enhanced) IMF as shown in \citetalias{labarbera2019}. For this reason, we later include IMF-sensitive indices in our analysis to ensure that our results are self-consistent. We  note that the highest metallicity of both SSP model grids is not enough to match the value of the central bin, which requires [M/H] > 0.22.

\begin{figure}
	\includegraphics[width=\columnwidth]{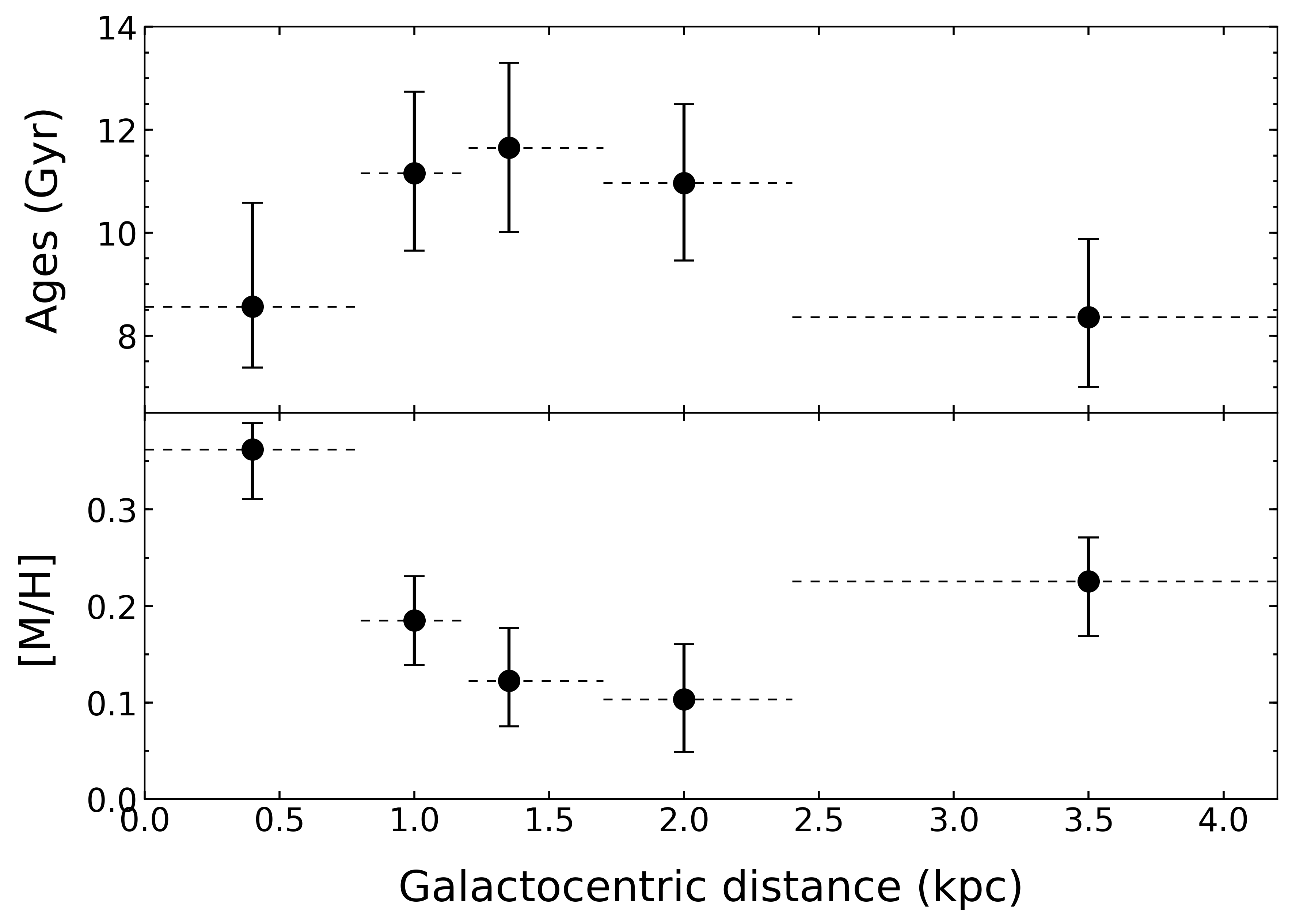}
    \caption{Mean luminosity-weighted ages and metallicities of each radial bin as a function of the distance to galaxy centre, derived by fitting the observed optical H$_{\beta_{o}}$ and [MgFe]' spectral indices with SSP model predictions. Horizontal dashed lines indicate the galactocentric distance range covered by each bin.}
    \label{fig:mlwa}
\end{figure}

We use these two indices to infer the mean luminosity weighted age (MLWA) and metallicity for each radial stacked spectrum, which are shown in Fig. \ref{fig:mlwa}. Here we assume a different IMF slope for each bin according to \citetalias{labarbera2019} (see Table \ref{tab:indexset}). Our derived stellar population parameters are in general agreement with the values derived in \citetalias{labarbera2019} for these galaxies. The age-sensitive index H$_{\beta_{o}}$ reveals very old stellar ages indicating that the bulk of stars in those regions were formed early. The age gradient is very shallow, in agreement with \cite{brough2007} and \cite{loubser2009} from optical spectroscopy analysis of BCGs. We note that the median MLWA is 11.0 $\pm$ 1.2 Gyr (the uncertainty comes from the standard deviation of the MLWAs). We derive a steep negative metallicity gradient with all values showing super-solar metallicities. This metallicity trend is in good agreement with what previous studies have reported for very massive galaxies (\citealt{kuntschner2010}, \citealt{labarbera2012}, \citealt{gonzalezdelgado2015}, \citealt{goddard2017}, \citealt{martinnavarro2018},  \citealt{ferreras2019}, \citealt{zibetti2020}).

\begin{figure*}
	\includegraphics[width=0.85\textwidth]{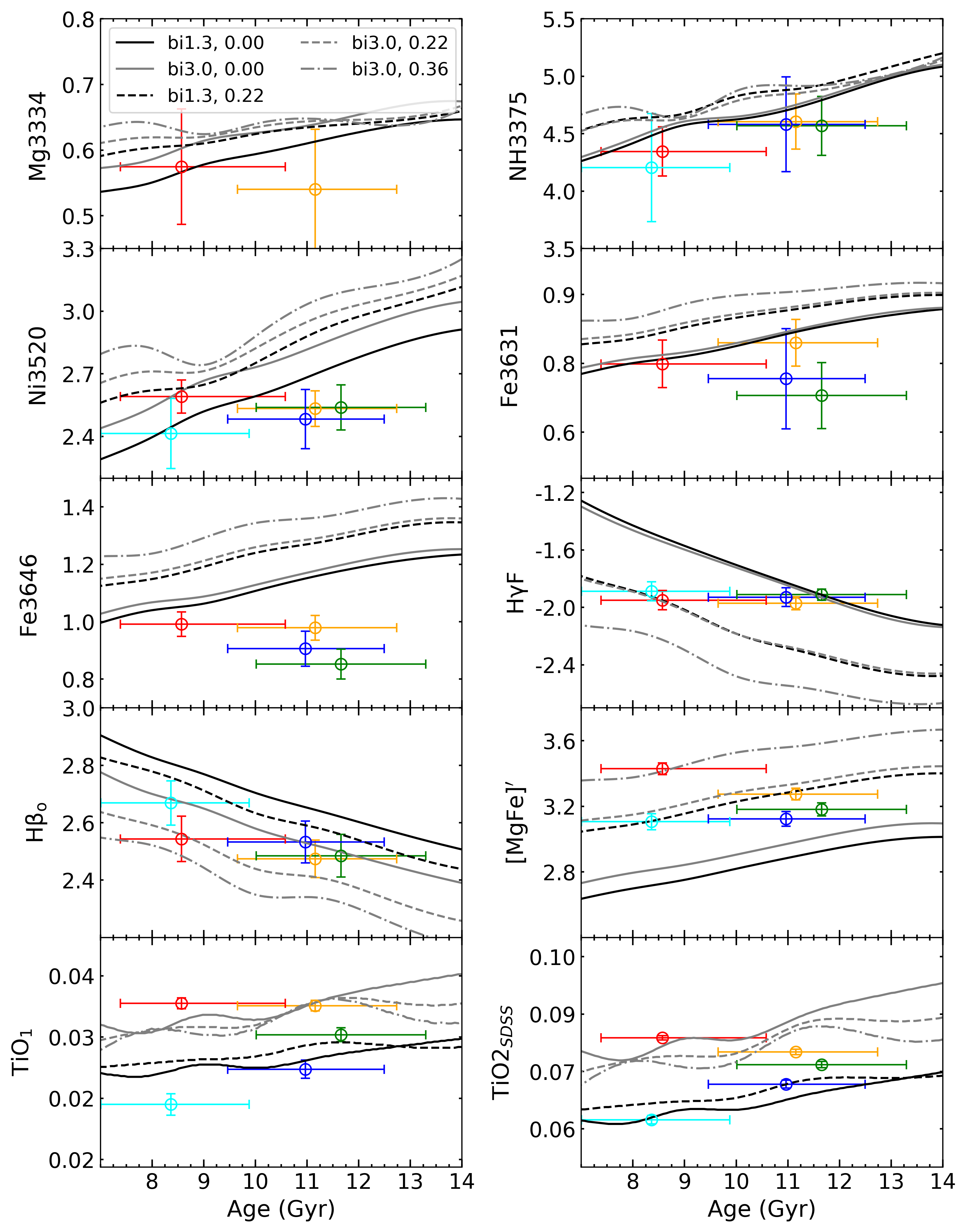}
    \caption{Observed and predicted SSP line-strengths are shown as a function of age for three different metallicities [M/H] = 0.0 (solid line),  [M/H] = 0.22 (dashed line) and [M/H] = 0.36 (dot-dashed line), and two IMFs, bimodal $\Gamma_{b}$=1.3 in black and a bottom heavy bimodal IMF $\Gamma_{b}$= 3.0 in grey. Note that predictions for [M/H] = 0.36 come from an extrapolation of the SSP models (see the text) and are plotted only for $\Gamma_{b}$= 3.0, to represent the innermost bin. Observed indices are overplotted as a function of their MLWA and color-coded as in Fig. \ref{fig:spectra}. Both SSP models and data are at the same spectral resolution 350 kms$^{-1}$.}
    \label{fig:indicesage}
\end{figure*}

\subsection{Observed versus model indices} \label{sec:indicesage}

In order to characterise the behaviour of the NUV and optical indices, we show the dependence of the SSP model predictions for each index with respect to the stellar population age, metallicity and IMF slope in Fig. \ref{fig:indicesage}. Measured indices at varying radial bins are shown with different colours. The lines represent the SSP model predictions as a function of the age for a standard ($\Gamma_{b}$=1.3, black) and bottom-heavy ($\Gamma_{b}$=3.0, grey) IMFs, and for three different metallicities: solar ([M/H] = 0.0, solid line), highest metallicity value from SSP models ([M/H] = 0.22, dashed line) and extrapolated metallicity ([M/H] = 0.36, dot-dashed line). The latter is plotted in order to represent the innermost bin for which we find a mean luminosity-weighted metallicity of [M/H]=0.36. Some indices show strong sensitivity to the age, such as the Balmer lines H$_{\beta_{o}}$ or H$\gamma$F, whereas, as expected, the [MgFe]' indicator is more sensitive to the metallicity. Note the strong IMF dependence of the TiO1 and TiO2$\rm_{SDSS}$ indices. The most remarkable aspect for the present work is that the galaxy indices in the NUV range depart significantly from the old SSP model predictions, where data fall below the old SSP model values. This indicates that a single old SSP is not a sufficient representation of the observed NUV indices (see also \citetalias{SR19} for a detailed discussion). These indices require an additional component to be matched. Combining NUV features with optical indices allows us to discriminate between a single old SSP and a more complex SFH, since the NUV range is biased towards the presence of small fractions of young stars. Here we aim to constrain these fractions from the departure of the NUV indices with respect to old SSP index values.

\subsection{SFH parametrization} \label{sec:sfhparam}

In this Section we describe our modelling of the SFH of a very massive galaxy to derive the young stellar component. As previously mentioned, massive galaxies form the bulk of their stars early over a relatively short period of time, with narrower SFHs that peak at older ages than their less massive counterparts which have more extended star formation periods (\citealt{vazdekis1997}, \citealt{ferreras2000}, \citealt{thomas2005}, \citealt{delarosa2011}, \citealt{mcdermid2015}). Many studies have derived the stellar population parameters in massive early-type cluster galaxies by approximating their SFHs with simplistic modelling approaches (\citealt{ferreras2000}, \citealt{pipino2009}, \citealt{groenewald2014}, \citealt{loubser2016}). Even a single burst at high redshift (i.e., an old SSP) provides a good approximation of their SFHs (\cite{trager2000}), at least judging from their optical spectra, for galaxies whose stellar mass is larger than $\rm 10^{11}~M_\odot$ (\citealt{vazdekis1997}, \citealt{renzini2006}), such as those studied here. However, Fig. \ref{fig:indicesage} clearly shows the inability of a unique and old SSP to match the observed NUV indices. A single SSP, while matching the optical indices, provides an unsatisfactory match to the observed NUV indices and suggests a more complex SFH with more than one star formation burst. The departure of the observed values with respect to an old SSP needs to come from a very hot stellar population \citetalias{SR19}. We interpret this departure as being due to the presence of young stars in these massive central galaxies, as previous studies have found recent star formation in some samples of BCGs. Our aim here is to characterize the young stellar component for each radial bin. If any other hot stellar components are contributing to this departure of NUV indices, such as the post-asymptotic giant branch stars, these are  sub-dominant contributions as has been discussed in \citetalias{SR19} and as is predicted by the stellar evolution theory (\citealt{buzzoni1989}, \citealt{charlot1991}, \citealt{conroy2013}).

To characterize this young component, we follow the modelling approach employed in \citetalias{SR19}. We model the SFH of a massive galaxy assuming a composite model described by two stellar components: an old and a young stellar population. The old stellar component is described by an SSP, which corresponds a single burst at high redshift that represents the bulk of the stars (see Section \ref{sec:mlwa}). For this stellar component, we assume prior distributions of ages from 1 to 13 Gyr and metallicities from --0.40 to 0.45 dex. The second component is parametrized by assuming a constant star formation rate (SFR) in the last 1 Gyr to estimate the relative fraction of stars formed during that period. The age and metallicity of the old component, the IMF slope and the mass fraction of young stars are left as free parameters in the fitting process. For each stacked spectrum, we compare the observed indices to index predictions from this modelling approach.

Note that we adopted 2 Gyr in \citetalias{SR19}, where we employed bluer indices with greater sensitivities to the young stellar components. Our present choice is aimed at maximising the sensitivity to the youngest stellar components. We note that our approach minimizes the effect of the burst age –- burst strength degeneracy, as the age of the young component is not a free parameter in the fitting process. Moreover, in Section \ref{sec:sr19comparison} we also show results when extending the upper age limit of the young component to 2 (rather than 1) Gyr to compare with our previous work \citetalias{SR19}, finding consistent results.


The young component may be represented by different SFH parameterizations. Certainly, one infers different young mass fractions depending on the parameterization of the SFH, but the main goal of this work is to assess the spatial distribution of young stars in massive central galaxies by constraining all the contributions within the last Gyr. If one considers a two-stellar population (2SSPs) model, i.e., one old stellar burst plus one young stellar burst, this might be representative for some specific massive galaxies, where they formed the bulk of their stars in a high-redshift star formation burst, and some little accretion later may produce a second burst at lower redshift, forming new stars in the centre. However, this secondary star forming epoch would be very unlikely to be the same for different galaxies, as the environment would randomly determine this star formation process. Therefore, since in this work we are averaging the spectra of six different galaxies, a constant SFH representing the young stellar component goes a step further with respect to the 2SSPs, in order to account for the stars formed in a given period of time (1 Gyr in this work) in our galaxy sample.

\subsection{Index fitting methodology}

The best-fit parameters for each radial bin are estimated by comparing the observed line-strengths indices with those predicted by the combined model described above. Our index fitting method is based on a Markov Chain Monte Carlo (MCMC) algorithm, using the publicly available \textsf{emcee} routine \citep{foremanmackey2013} to fit simultaneously the set of indices listed at Table \ref{tab:indexset} for each stacked spectrum. The best-fitting model is obtained by maximizing the log-likelihood function $\ln\mathcal{L}$. That is, given an observed set of indices ($\mathcal{I}_{stack}$), the likelihood is equal to the probability $\mathcal{P}$ of the model parameters ($\Theta$ = [Age$\rm_{old}$, [M/H]$\rm_{old}$, $\Gamma_{b}$, f$\rm_{young}$]) of getting the observed indices ($\mathcal{I}_{stack}$):

\begin{equation}
\mathcal{P}(\mathcal{I}|\Theta) = \ln\mathcal{L}(\Theta|\mathcal{I}_{stack}),
\end{equation}
where the likelihood is defined by $\ln\mathcal{L}(\Theta|\mathcal{I}_{stack}) \propto - \frac{\chi^{2}}{2}$, where the $\chi^{2}$ statistic is:
\begin{equation}
\chi^{2} =   \sum_{i}^{N} \left(\frac{I_{\rm model_{i}}-I_{\rm stack_{i}}}{\sigma_{\rm stack_{i}}}\right)^2,
\end{equation}
where the subscript $i$ refers to the $i$-th line index, $I_{\rm model}$ and $I_{\rm stack}$ are the model and observed indices, respectively, and $\sigma_{\rm stack}$ represents the uncertainty on the indices.

The model that maximizes the likelihood is the best-fit model. Each set of indices measured in each radial bin is fitted with the index predictions from the SFH parametrization described in Section \ref{sec:sfhparam}. We give equal relative weight to the optical and NUV indices, separating the $\chi^{2}$ statistic at 3700 \AA, slightly rescaling the error bars so that each set of indices has the same weight on $\chi^{2}$.

Each MCMC simulation consist of 40 walkers exploring the parameter space over 40000 steps. This process performs a linear interpolation of the SSP models predictions to generate continuous coverage of the parameter space given by the age and metallicity of the old stellar component, and the fraction of young stars of each model template. We fit the observed line-strengths of each radial bin to obtain the best-fit model parameters via MCMC chains. The best fit values are the median of each chain distribution parameter and the uncertainties in the median are the 16-th and 84-th percentile levels, which are marginalized over the other parameters of our modelling approach. For the IMF we do not perform a continuous coverage and, instead, we use SSP models with IMF slopes {1.0, 1.3, 1.5, 1.8, 2.0, 2.3, 2.5, 2.8, 3.0, 3.3, 3.5} and obtain the best-fitted IMF via minimization of the $\chi^{2}$. This process gives the best-fit values for the set of model parameters of each stacked spectrum: the age and metallicity of the old stellar component, Age$\rm_{old}$ and [M/H]$\rm_{old}$, respectively, the mass fraction of the stars formed in the last 1 Gyr f$\rm_{young}$ and the IMF slope. The results are further discussed in Section \ref{sec:results}.

\section{Results} \label{sec:results}

In this section we present the stellar population parameters obtained with the method described above. Fig. \ref{fig:indicesage} shows the mismatch between the observed and the model NUV indices for a single, old SSP, which provides good fits to the optical indices. This mismatch can be resolved by adding a small amount of recent star formation on top of an old stellar population. We derive the young stellar population mass fraction in the last 1 Gyr for each radial stacked spectrum. We also infer this fraction for the innermost bin of each individual galaxy to detect possible differences among the six galaxies.

 
\subsection{Young stellar population gradients}

\begin{figure}
	\includegraphics[width=\columnwidth]{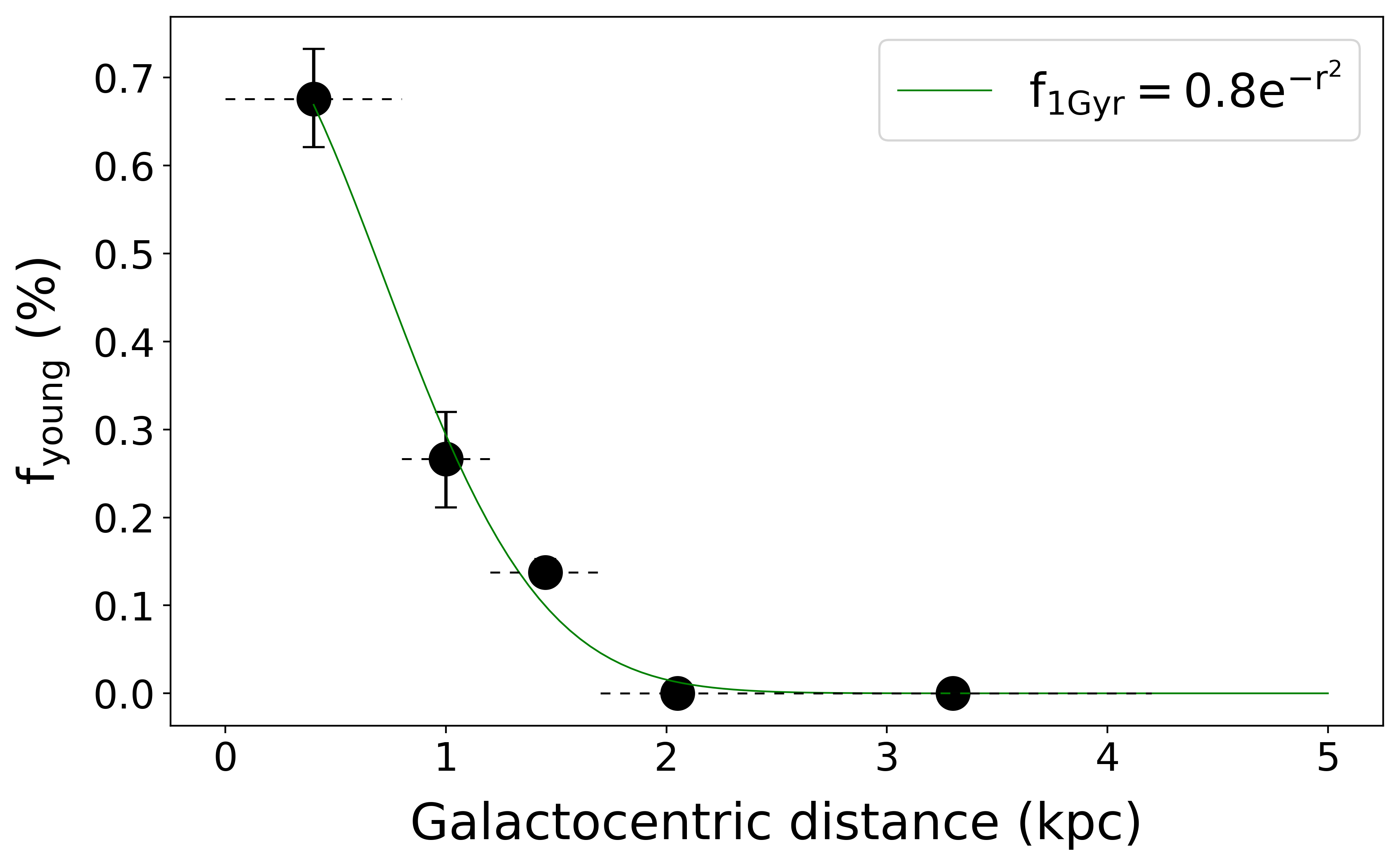}
    \caption{Variation of young mass fractions with galactocentric radius. The mass fractions of stars formed in the last 1 Gyr (f$\rm_{young}$) are plotted as a function of the galactocentric distance in units of kpc. Black symbols are the best-fitting values from the index fitting analysis. Horizontal dashed lines indicate the galactocentric region covered by each bin. Green solid line indicates the polynomial fit (in the inset) to the best-fitted values up to 5 kpc.}
    \label{fig:fractionsfit}
\end{figure}

Results for the young stellar component for each radial stacked spectrum are provided in Fig. \ref{fig:fractionsfit}. Best-fit values for the other fitted parameters are summarized in Table \ref{tab:bestfit}. We measure mass fractions for the young stellar component f$\rm_{young}$ below 1\% at all galactocentric distances. There is a larger contribution of young stellar populations in the innermost regions with a mass fraction of $\sim$0.7\% in the central bin decreasing to $\sim$0.3\% at 1 kpc distance. Beyond that, the gradient becomes shallower with virtually no contribution from young stars outwards of 2 kpc. Therefore, the young component is mostly concentrated in the galaxy cores (<2 kpc) of our sample of massive central cluster galaxies. 

\begin{table}
 \centering
 \begin{tabular}{ccccc}
 \hline	
Bin (kpc) & IMF & Age$\rm_{old}$ (Gyr)  & [M/H]$\rm _{old}$ & f$\rm_{young}$ (\%)\\
  (1) & (2) & (3) & (4) & (5) \\
 \hline
0.0 -- 0.8  & 3.0 & 11.6$^{+0.2}_{-0.3}$ & 0.32$\pm0.01$ & 0.67$^{+0.06}_{-0.05}$\\
0.8 -- 1.2 & 2.8 & 12.0$^{+0.4}_{-0.3}$ & 0.19$\pm0.02$  & 0.27$\pm0.06$\\
1.2 -- 1.7 & 2.5 & 12.2$^{+0.2}_{-0.3}$ & 0.15$\pm0.01$ & 0.14$^{+0.02}_{-0.01}$ \\
1.7 -- 2.4 & 1.8 & 11.4$^{+0.7}_{-0.5}$ & 0.14$\pm0.03$  & 0 \\
2.4 -- 4.2 & 1.3 & 9.0$^{+2.0}_{-1.0}$ & 0.20$\pm0.05$ & 0 \\
 \hline
 \end{tabular}
\caption{Best-fit values of the stellar population parameters. For each radial bin listed in column 1 we have obtained the best-fit IMF slope (column 2), and age and metallicity for the old stellar component (columns 3 and 4, respectively). The young fraction is shown in column 5. Note that the metallicity of the old component in the first bin exceeds the upper prior [M/H]$\rm _{old}$=0.22 of the E-MILES SSP models. 
   }
 \label{tab:bestfit}
\end{table}

All the parameter estimates are listed in Table \ref{tab:bestfit}. The old component shows metal-rich metallicities decreasing with galactocentric distance, in agreement with previous studies for massive ETGs. Note that the metallicity for the innermost bin is [M/H] = 0.30, which comes from a small linear extrapolation of the SSP models. The best-fit metallicity of the innermost bin before the extrapolation was the highest value allowed in the E-MILES models, i.e., [M/H] = 0.22, giving f$\rm_{young}$=0.42\%. This extrapolation to higher metallicities improved the fits to constrain properly the young component for this radial bin. In the case of the two outermost bins, the best-fit f$\rm_{young}$ of our modelling approach is zero, i.e., consistent with a single, old SSP. Hence the age and metallicity derived in Fig. \ref{fig:mlwa} are a good representation of the stellar population in these bins.

We find that by including the IMF-sensitive TiO spectral indices (\citealt{labarbera2013}, \citealt{martinnavarro2018}, \citealt{eftekhari2019}), we infer IMF values as a function of radius consistent with those derived in \citetalias{labarbera2019} for the same set of galaxies, which is based on a larger set of IMF-sensitive indices. This variation of the IMF with radius in massive ETGs is in agreement with previous studies (\citealt{martinnavarro2015}, \citealt{labarbera2016}, \citealt{vaughan2018}, \citealt{sarzi2018}, \citealt{meyer2019}, \citealt{labarbera2019}). 

\begin{figure*}
	\includegraphics[width=0.85\textwidth]{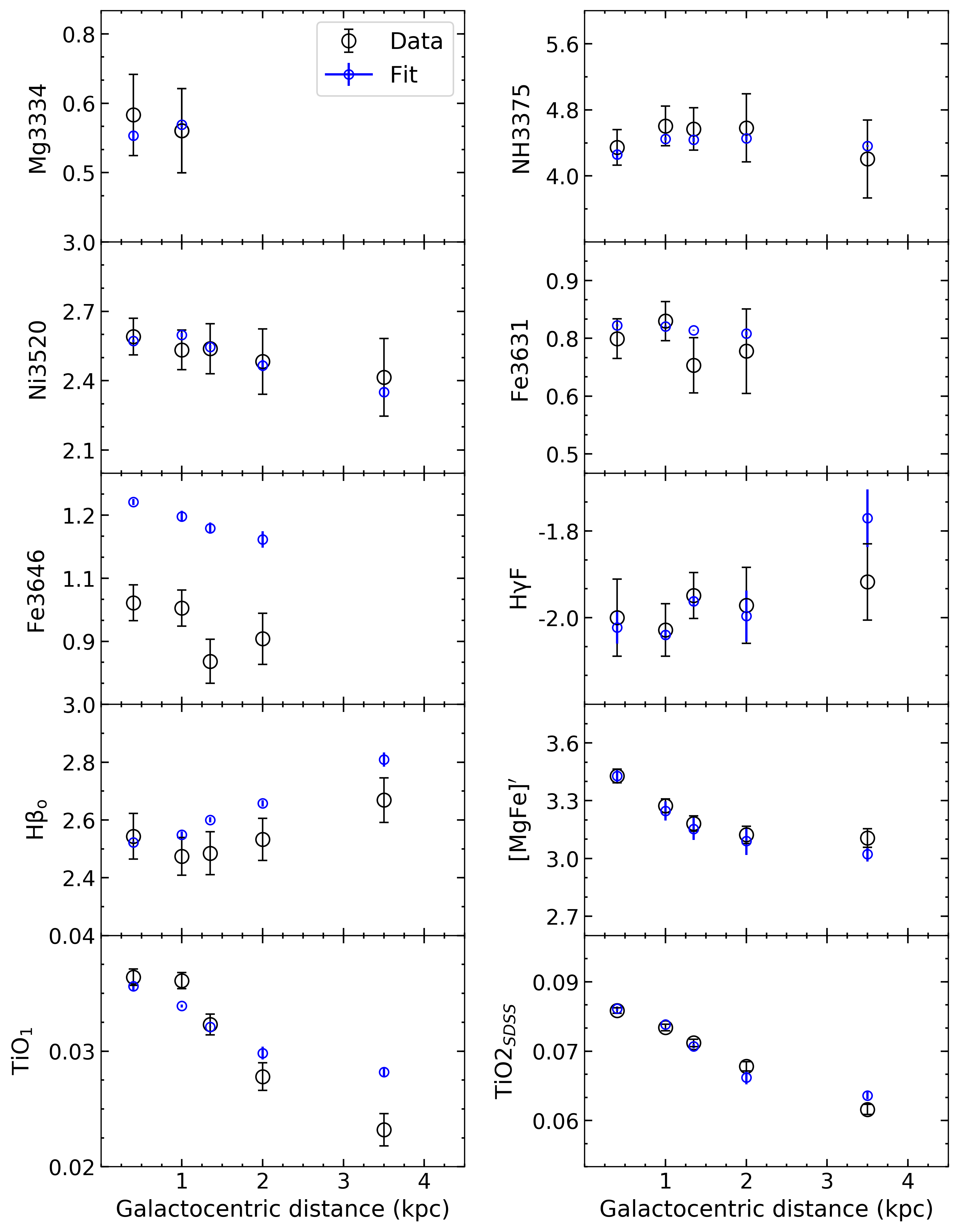}
    \caption{Best-fit spectral indices as a function of the galactocentric distance of each radial bin are plotted in blue, and the measurements on the observed spectra are shown in black. Note that different sets of indices are used at each radial bin during the fitting process to obtain the best-fit parameters indicated in Table \ref{tab:bestfit}.}
    \label{fig:indicesradius}
\end{figure*}

We show in Fig. \ref{fig:indicesradius} the observed (black) and best-fitting (blue) model indices as a function of galactocentric distance. Some indices show a clear trend with distance to galaxy centre, such as the negative gradients for the Fe and Ni NUV indices, and the optical [MgFe]' and the TiO indices, as expected. Balmer lines present slightly positive gradients. The best-fits show very good agreement with the observed indices. The NUV indices are well fitted, except for the Fe3646 line. This could be due to the fact that this index is still affected by alpha-element abundance, despite the fact that we have selected indices that are more sensitive to small fractions of young stars. However, the alpha-enhancement may produce a larger effect on this index that we are not taking into account with the SSP models used, than in the other NUV indices, since the impact of alpha-elements on SSP models is more uncertain in the UV than is the case for the optical. Therefore, both effects could be contributing to the Fe3646 index, but in this work we are only constraining the young stellar component. We have checked that the exclusion of this index does not affect the results. The optical best-fitting indices are in agreement with the observations.


\subsection{Uncertainties} \label{sec:uncertainties}

\begin{figure}
	\includegraphics[width=\columnwidth]{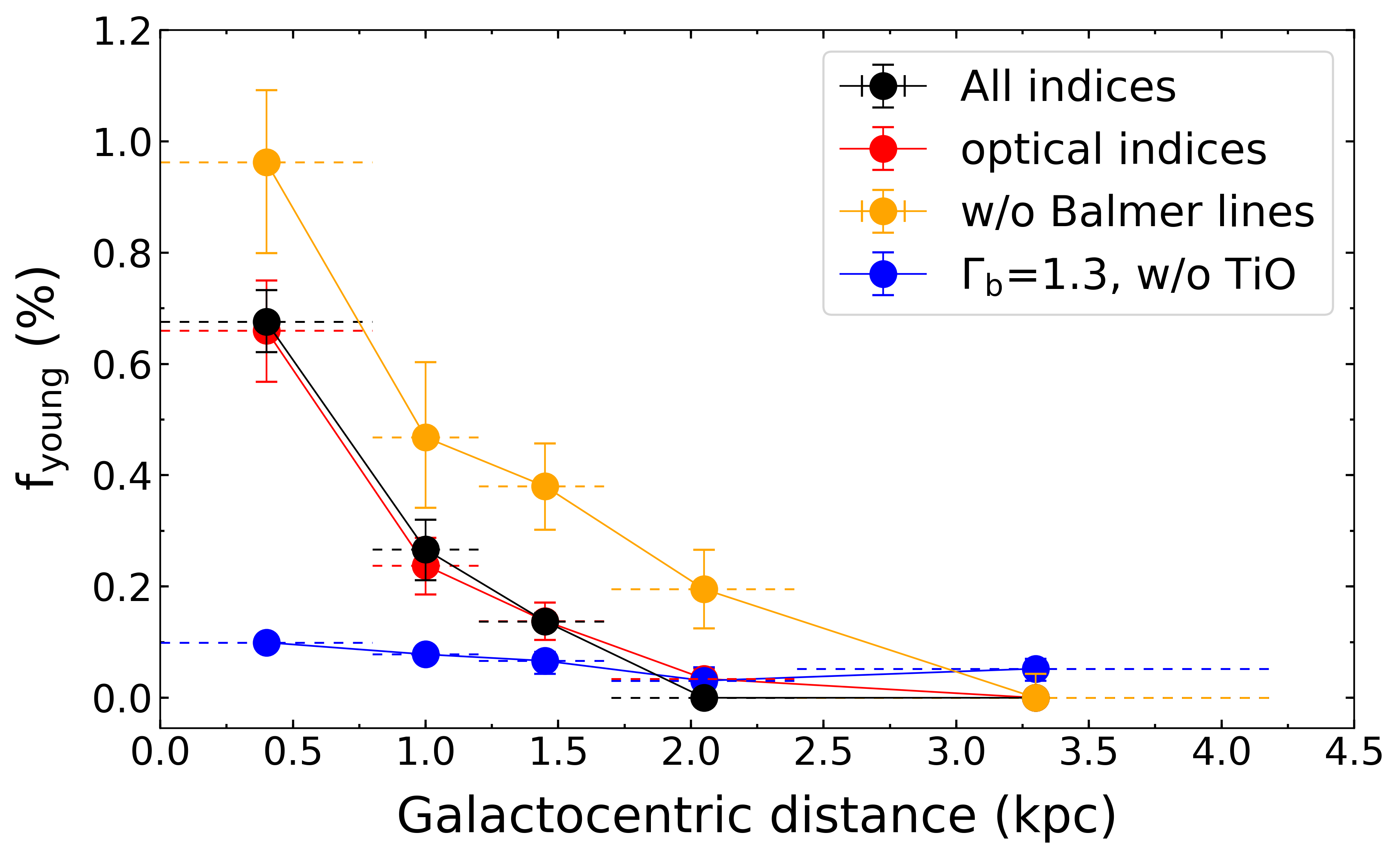}
    \caption{The panel shows the fractions of young stars when using the full set of indices listed in Table \ref{tab:indexset} (black) and when excluding, from the set of indices used in the fitting process, the optical age-sensitive Balmer lines (orange) and the NUV indices (red). Also shown are the results when fixing the IMF as $\rm\Gamma_{b}\sim$1.3 for all the radial bins (blue) and removing the TiO IMF-sensitive indices in the fitting process.}
    \label{fig:fractionsindices}
\end{figure}

The derived fractions of young stars are only meaningful if accompanied with a comprehensive analysis of the errors associated to the obtained fractions. As the Balmer lines are the main age indicators in the optical range we investigate the impact of removing these indices from the fitting process on the derived fractions. In Fig. \ref{fig:fractionsindices} we see that the removal of the Balmer lines lead to higher young mass fractions at all the radial bins with respect to those obtained with all the index set. This is not surprising given the fact that in this case the age constraints come from the NUV indices, which are mostly contributed by the youngest populations. However, the trend with galactocentric distance is maintained. Small fractions of young components in massive ETGs can be derived also by using extremely high S/N spectra in the optical range (\citealt{loubser2009}, \citealt{martinnavarro2018}, \citetalias{SR19}). If only fitting the optical indices, we obtain rather similar results as shown in the figure, albeit with slightly lower young star fractions and larger uncertainties. The agreement found in the solutions between the optical and the optical plus the NUV indices shows that our results are very robust, as we are getting fully consistent results across the UV and optical spectral ranges.

We also investigated how the derived parameters are affected when using a standard bimodal IMF slope of $\Gamma_{b}\sim$1.3 for all radial bins. This is a conservative approach that departs from what previous studies have found: an IMF gradient from bottom-heavy IMF slope in the central regions of massive galaxies and a rapid decrease to Milky Way like IMF at larger radii. As the TiO indices cannot be fit with a standard IMF we have excluded TiO1 and TiO2$\rm_{SDSS}$ in the fitting process, thus avoiding biasing significantly the results. This is particularly evident for the two innermost bins where the IMF deviates significantly from a standard distribution. Fig. \ref{fig:fractionsindices} shows smaller fractions of young stars (f$\rm_{young}$<0.1\%) with a shallower trend with galactocentric distance in the case of $\Gamma_{b}\sim$1.3, than when the IMF is set as a free parameter. This is an indirect effect of the larger age estimate obtained for the old stellar component that, by being fainter, it requires a smaller young fraction to provide the observed departure in the NUV indices. Note that in Fig. \ref{fig:grid}, the two innermost bins have ages older than 13 Gyr if $\Gamma_{b}\sim$1.3, which are older than when considering a bottom-heavy IMF. Therefore, the effect of the IMF is to change the radial gradient of the age of the old component and, as a consequence, the gradient of the fractions of young stars. This highlights the importance of considering the IMF by including IMF-sensitive indices in the analysis for this type of extremely massive galaxies. This is particularly relevant for their galaxy cores, which are characterised by bottom-heavy IMFs.


We also note that we have not corrected the spectra for internal dust extinction in these galaxies; the shape of the continuum in our spectra may be affected by reddening, specially to bluer wavelengths. However, dust will mostly affect measurements covering large wavelength intervals, e.g. broadband photometry, and much less the narrow spectral features used in this work. Therefore, the effect of dust in the analysis of the spectral indices studied is negligible for the indices used in this study (\citealt{macarthur2005}, \citetalias{SR19}). Additionally, our targets have been selected to have low internal extinction, as detailed in \cite{labarbera2017} and  \citetalias{labarbera2019}.

\subsection{Young stellar populations in the cores of individual galaxies} \label{sec:indivspectra}

We also determine the young mass fractions in the innermost regions for each individual galaxy in order to detect possible variations among the six galaxy cores. This can only be achieved for the innermost individual galaxy spectra (0 -- 0.8 kpc) due to their sufficiently high S/N in the NUV spectral range. For each spectrum, H$_{\beta_{o}}$ line is corrected for nebular emission contamination as detailed in \citetalias{labarbera2019}. The contamination is significant only for XSG1, XSG6 and XSG8 galaxies. We consider a bottom-heavy IMF slope for each galaxy according to the \citetalias{labarbera2019} estimates. The derived parameters are presented in Table \ref{tab:centralspectra}. All our galaxies present signs of young stars in their central regions. The stellar populations younger than 1 Gyr contribute less than one percent in all galaxies. The median value for our sample of 6 BCGs is 0.70$\pm{0.24}$\%, in agreement with our estimate of the innermost stacked spectrum (Table \ref{tab:bestfit}). Note that the metallicity of the old component for all galaxies is higher than the maximum value allowed by the E-MILES models ([M/H]=0.22), so we have performed a linear extrapolation of the metallicity up to [M/H]=0.45.

As explained in Appendix~A of \citetalias{labarbera2019}, XSG1, XSG6, XSG8, XSG9, and XSG10 are central group galaxies, according to the updated SDSS-DR7 group catalogue of \cite{wang2014}. XSG7 shares the central position of a galaxy group with another galaxy of similar luminosity. Also, XSG1, XSG6 and XSG8 have more prominent emission lines. In particular XSG6 is the galaxy with strongest emission in this work, being apparent at all the galactocentric distances. Based on these results, if one assumes that environment enhances the levels of star formation, one may expect galaxies like XSG6 to have the highest fractions of young stars, while XSG2 -- not located at its cluster galaxy core -- to have lowest f$\rm_{young}$ (see App. \ref{app:xsg2}). However, we obtain that the young stellar contribution in XSG6 is the lowest. Moreover, we can not infer properties about the link between the young stellar populations in massive galaxies and their surrounding environment. Although it is remarkable that our BCGs show rather similar young mass fractions being smaller than 1\%, with the exception of XSG6 for which we obtain less than 0.2\%, we are aware that our sample is small.

\begin{table}
 \centering
 \begin{tabular}{lccc}
 \hline	
ID & Age$\rm_{old}$ (Gyr) & [M/H]$\rm _{old}$ &f$\rm_{young}$  \\
  (1) & (2) & (3) & (4) \\
 \hline
XSG1 & 10.88$\pm{0.06}$ & 0.28$\pm{0.01}$ & 0.77$\pm{0.02}$ \\
XSG6 & 10.73$\pm{0.06}$ & 0.25$\pm{0.01}$ & 0.14$\pm{0.03}$ \\ 
XSG7 & 11.75$^{+0.05}_{-0.02}$ & 0.28$\pm{0.01}$ & 0.84$^{+0.04}_{-0.02}$ \\
XSG8 & 10.63$\pm{0.05}$ & 0.41$\pm{0.01}$ & 0.62$\pm{0.03}$ \\
XSG9 & 10.2$\pm{0.1}$ & 0.36$\pm{0.01}$	& 0.48$\pm{0.03}$  \\
XSG10 & >13.0 & 0.24$\pm{0.01}$ & 0.77$\pm{0.02}$ \\
 \hline
 \end{tabular}
\caption{Best-fit values of the stellar population parameters for the individual galaxy spectra at the innermost regions (<0.8 kpc). For each galaxy listed in column 1 we obtain the best-fit ages and metallicities for the old component reported in columns 2 and 3, respectively. Column 4 reports the best-fit fraction of young stars of each galaxy. 
   }
 \label{tab:centralspectra}
\end{table}

\subsection{Comparison with the young mass fractions obtained for massive galaxies at z$\sim$0.4} \label{sec:sr19comparison}

In this section we compare our results with the fraction of young stars in normal ETGs (\citetalias{SR19}) for which we employed a similar methodology. \citetalias{SR19} analysed stacked spectra from thousands of very massive ETG spectra from SDSS (with central velocity dispersion ranging from 220 -- 340 kms$^{-1}$), at average redshift z$\sim$0.4. At this redshift, the aperture  $\rm \phi_{R}$ of the SDSS fibers corresponds to a radius of R = 5 kpc. From the combination of line-strength indices in the spectral window 2500 -- 6000 \AA, \citetalias{SR19} found that massive ETGs at z$\sim$0.4 are home to low levels of recent star formation. For ETGs whose central velocity dispersion is larger than 300 kms$^{-1}$ -- similar to our galaxies --, they derived mass fractions of stars formed in the last 2 Gyr of $\sim$0.48$\pm$0.01\%. To perform a fair comparison with this study, it is necessary to compute the young mass fractions in the same way as described in Section \ref{sec:sfhparam} but, in this case, for a 2 Gyr instead of 1 Gyr period of the young stellar component and integrate these fractions  within an aperture $\rm \phi_{R}$ = 5 kpc.

We need to know the stellar mass of the young component ($\rm M_{young}$) within that radius. Since the IMF slope is known at each radial distance, we follow the equation written in Fig. 6 from \citetalias{labarbera2019} to derive the stellar mass density $\Sigma$ as a function of radius up to 5 kpc from the IMF slope ($\Gamma$), written as:

\begin{equation}
\Sigma = -0.24 \rm ln\left( -1 + \frac{1.84}{\Gamma-1.3}\right) 
\label{eq:stellarmass}
\end{equation}

The stellar mass density is multiplied by the area covered in each radial bin to derive the total stellar mass ($\rm M_{\star}$) in each stack. Then, we fit an exponential function to the young mass fractions to extrapolate them out to 5 kpc. We multiply those fractions by the total stellar mass derived before, in order to obtain $\rm M_{young}$ as a function of radius. Finally, the fraction of young stars within a given aperture f$\rm _{\phi_{R}}$ can be computed with:

\begin{equation}
\rm f_{\phi_{R}} =  \frac{\sum^{r=R}_{r=0kpc} M_{young}}{\sum^{r=R}_{r=0kpc} M_{\star}}
\label{eq:fractionradius}
\end{equation}

We obtain a f$_{\phi_{R}}\sim$0.42\% of young stars within an aperture of $\rm \phi_{R}$ = 5 kpc. Interestingly, the mass fraction of young stars within 5 kpc is slightly lower than the fraction derived in \citetalias{SR19}. Note that our objects are nearby galaxies located at the centre of a galaxy cluster or group. From abundance matching methods, \citetalias{SR19} concludes that at least half of their galaxies are located in galaxy clusters with mass larger than 10$^{14}$ M$_{\odot}$. However, whether they are central or satellite galaxies remains unknown. Therefore, we need to have in mind that we are not comparing exactly the same class of objects, but very massive galaxies at different redshifts with an early-type morphology and very high central velocity dispersions. If we assume that our objects are similar to the galaxies in the most massive bin (central velocity dispersion of 300 -- 340 kms$^{-1}$) of \citetalias{SR19} but at different epochs we may conclude that these lower young mass fractions within the central 5 kpc regions are consistent with passive evolution. Lower contributions of young stars are expected at lower redshifts as a consequence of the decrease of gas available to form new stars. Moreover, if the required gas is contributed by mergers, it would be also expected to decrease at lower redshift, similar to the expected merger rate (\citealt{lotz2011}, \citealt{lopezsanjuan2015}).


\section{Summary and conclusions } \label{sec:discussion}

Central galaxies in clusters are very special places to study galaxy formation and evolution. Line-strength indices of their integrated spectra provide important insights on their stellar populations. In this paper, we analysed the average radial distribution of the young stellar component of six massive BCGs. We divide the individual galaxy spectra in five radial bins and stack them to obtain representative spectra with high S/N. The set of NUV indices have been selected so that they are highly sensitive to young stars, and the optical indices sensitive to the bulk of the stellar population. The latter include indices sensitive to the age, metallicity and IMF slope. We show that the measured NUV indices in our stacked spectra are not well represented by a single, old SSP model, as is the case if only the optical indices are fitted, since they show a departure from the old SSP index values. We interpret this departure as evidence for a young stellar component. We investigated the young stellar gradients with an index-fitting approach, by comparing the observed indices with the model predictions that come from a simple model of the SFH of a massive galaxy. This model consists of a superposition of a single burst representing the old stellar component, and a young component that is characterized by assuming constant SFR in the last 1 Gyr. The main results are summarized below.

From H$_{\beta_{o}}$ and [MgFe]' we measured the mean luminosity-weighted ages, metallicities and  IMF slopes of each radial bin. We find old ages with an average MLWA of 11.8 Gyr. The best-fit metallicities present a negative gradient with super-solar metallicity in all radial bins. The stacked NUV index values show a mismatch with respect to the old stellar population model predictions of their MLWAs, suggesting that a SSP is not the optimal representation so that an additional component is required, which is mainly contributing to bluer wavelengths. This discrepancy between the predicted and the measured NUV indices can be addressed by adding small fractions of young stellar populations. 

Thanks to the highly IMF-sensitive indices included in the set of indices, TiO1 and TiO2$\rm_{SDSS}$, we have been able to fit the IMF slope. The obtained values are in agreement with those derived by \citetalias{labarbera2019}. As it has already been studied by previous authors, there is a clear IMF variation with radius, from bottom heavy at the central regions to a standard IMF at larger radii. We find that it is crucial to account for the radially varying IMF in massive galaxies in order to correctly determine the young stellar gradients via the NUV.

The key finding in this study is a negative gradient of the young stellar contribution in our stacked spectra of BCGs. The results show that  young stars (<1 Gyr) are localized in the cores of massive central cluster galaxies, where we measure a young mass fraction of f$\rm_{young}\sim$0.7\% at < 0.8 kpc. At larger radii this decreases rapidly to virtually an absence of young stars beyond 2 kpc. With these small components of young stars on top of a dominant old population we are able to match the observed indices of each radial bin as shown in Fig. \ref{fig:indicesradius}.

A 0.7\% fraction of young stars implies that the SFR in the last 1 Gyr of their SFHs (i.e. at redshift z<0.1) in the galaxy cores of our galaxy sample is, on average, 2.2 M$_{\odot}$yr$^{-1}$. Such a level of star formation activity is possible due to the existence of reservoirs of cold gas in their central regions. The source of this gas supply to form new stars remains unclear. It could be associated with intrinsic processes such as pristine gas not yet transformed into stars or recycled gas from stellar evolution, or extrinsic processes such as accreted gas-rich galaxies or the intercluster medium that reach the galaxy core. BCGs are thought to experience a large number of galaxy mergers and interactions with the satellite galaxies surrounding them. Numerical simulations have shown that the inner galaxy regions are dominated by in-situ stars, i.e., stars formed within the host galaxy, whereas the outer regions are mostly populated with ex-situ stars accreted during mergers with smaller galaxies (\citealt{oser2010}, \citealt{navarrogonzalez2013}; \citealt{shankar2013}, \citealt{cooper2015}). The accreted material dominates the outer galaxy regions, but whether the accreted gas reaches the central regions of central cluster galaxies and cools down to be able to form these new stars remains unknown. With the methodology and data used in this work we are not able to constrain the metallicity of the young component. This would give hints towards the origin of the gas employed to form such new stars.

The behaviour of the IMF places interesting constraints on the origin of this gas. The IMF regulates the amount of gas returned to the interstellar medium (ISM) due to stellar evolution. A bottom-heavy IMF is dominated by low-mass dwarf stars with longer lifetimes than the Hubble time, which lock-up most of this gas making it unavailable for star formation. For example, according to the E-MILES models, a Kroupa-like IMF $\Gamma_{b}=1.3$ returns a gas fraction of f$\rm_{gas}\sim$36.7\% into the ISM after 10.5 Gyr (i.e., the age of the old stellar population before the 1 Gyr period that we consider for constraining the young component). However, a stellar population with a bottom-heavy IMF $\Gamma_{b}=3.0$ returns an order of magnitude less gas: f$\rm_{gas}\sim$3.3\%. From eq. \ref{eq:stellarmass}, we infer that the stellar mass of the old stellar population in the innermost bin (<0.8 kpc) is 3.2$\times10^{11}$ M$_{\odot}$. Since for this bin $\Gamma_{b}=$3.0, this implies that the amount of gas returned to the ISM from this population is $\sim$1$\times10^{10}$ M$_{\odot}$. Interestingly, the stellar mass of the young stellar component in this region is 2.2$\times10^{9}$ M$_{\odot}$. I.e., this amount of gas could be responsible for producing the young stars we observe if this gas is able to cool. In this case, the star formation efficiency (SFE), i.e, the star formation rate per unit gas mass, would be 2.1 $\times10^{-10}$ yr$^{-1}$ within the last 1 Gyr. However, this SFE is one order of magnitude lower than the \cite{kokusho2017} estimates for nearby ETGs of similar stellar masses. Therefore, only 10\% of this gas would be required to cool down and form stars in order to match their SFE estimates. If the innermost bin, instead, were populated by a Kroupa-like IMF stellar population, as considered in Section \ref{sec:uncertainties}, the SFE would be 3 orders of magnitude lower than the \cite{kokusho2017} results and only 0.1\% of the gas returned into the ISM would be required to cool down to form new stars. Additionally, from the equation shown in Fig. 5 of \cite{martig2013}, which establishes a correlation between the SFR surface density and the gas surface density for a sample of ETGs and late-type galaxies, we infer a star formation rate surface density in good agreement with our estimates. Our results suggest that the young stellar populations found in the central regions of our BCGs may be the result of star formation fueled by gas returned from the evolution of the old stellar populations that dominate the bulk of these galaxies. However, due to the particular location of these objects in their host clusters, BCGs constantly interact with and are shaped by their surrounding environment. Several studies have observed a trend with the fraction of star-forming BCGs and halo mass, where this fraction decreases with the halo mass (\citealt{oliva2014}, \citealt{cerulo2019}). However, our galaxy sample is too small to study this trend with the halo mass accurately, and therefore a larger sample should be analysed.

Some residual gas might also be left over from a first phase of galaxy formation when the IMF was possibly top-heavy. This phase is required in order to match the chemical content of massive ETGs (\citealt{vazdekis1997}, \citealt{weidner2013}, \citealt{ferreras2015}, \citealt{jerabkova2018}). Therefore, the derived negative gradient of young stars in BCGs is consistent with an in-situ star formation activity. At the same time, the results of \citetalias{SR19} indicate that an ex-situ origin for the gas is unlikely due to the near constant SF fraction in BOSS galaxies (accreted gas might be expected to be more stochastic, i.e., show a greater variation of young mass fractions). Galaxy cores are where larger amounts of available gas are accumulated due to their deep gravitational potential wells, which are expected to retain gas more efficiently than in the outer regions. Moving outwards, the star formation activity produces increasingly smaller fractions of young stars, so that beyond 2 kpc no new stars are being formed. Note, however, that these results are average estimates for our sample of six BCGs. Given that the sample studied in this work is small, it is not possible to draw more statistically meaningful conclusions about the existence of young stellar population gradients in the BCG population as a whole. While this study is not able to answer which is the main mechanism that fuels the late star formation in BCGs, these young stellar gradients are in agreement with those works that associate the "blue-core" in some sample of BCGs to the presence of active star formation with young stellar mass fractions lower than 1\% (\citealt{bildfell2008}, \citealt{pipino2009}). More recently, \cite{cerulo2019} reported that the star formation in BCGs is more frequent at higher redshifts and decreases with cluster mass and galaxy stellar mass. Since more massive BCGs lie in the centre of more massive clusters the young stellar gradients should be further investigated by a larger sample of individual BCGs so we can determine whether the star formation is affected by the surrounding environment in each galaxy cluster. Very important constraints to the SFH of BCGs should also come from studying how the young stellar gradients of massive central galaxies evolve with redshift.

The analysis of the innermost regions of the individual BCGs, shows that all galaxies exhibit fractions of young stars smaller than 1\% within 0.8 kpc. However, given the small sample size, we are not able to derive general properties about the young contributions of the BCG population compared to those  from non central cluster galaxies. Further investigations should be performed along this line including a wider variety of environments of the most massive galaxies in the Universe.

\section*{Acknowledgements}


NSR acknowledges funding from Spanish Ministry of Science, Innovation and Universities (MCIU), through research project SEV-2015-0548-16-4 and predoctoral contract BES-2016-078409. MAB acknowledges support from the Severo Ochoa Excellence scheme (SEV-2015-0548). NSR, MAB and AV acknowledge support from grants AYA2016-77237-C3-1-P and PID2019-107427GB-C32 from the MCIU. FLB acknowledges financial support from the European Union’s Horizon 2020 research and innovation programme under the Marie Sklodowska-Curie grant agreement n. 721463 to the SUNDIAL ITN network.

\section*{Data availability} 

This work is based on observations made with ESO Telescopes at the Paranal Observatory under programmes ID 092.B0378, 094.B-0747, 097.B-0229 (PI: FLB). The stacked spectra data created during the current study that supports the plots within this paper are available from the corresponding author upon request. The E-MILES SSP models are publicly available at the MILES website (http://miles.iac.es).




\bibliographystyle{mnras}
\bibliography{references} 




\appendix

\section{Young stellar components of the satellite galaxy XSG2} \label{app:xsg2}

In this section we analyse the young stellar contribution of galaxy XSG2, which is the only target of the sample used in \citetalias{labarbera2019} classified as a satellite. Named J002819.3-001446.7 according to the SDSS, this galaxy is very similar to the other galaxies: a very massive early-type system with high central velocity dispersion. However, we exclude XSG2 from the main analysis of this work since it is the only galaxy that is not located in the centre of its parent cluster.

We have explored how the best-fit parameters differ when including the XSG2 galaxy in the stacked spectra. Results are shown in Table \ref{tab:xsg2}. Overall, we find similar stellar population properties. Among the differences we obtain f$\rm_{young}$=0.75\% for the innermost bin, which is slightly higher than the value obtained when this galaxy is excluded (see Table \ref{tab:bestfit}). Therefore, XSG2 has a slightly larger young stellar component in its centre compared to our main sample. The young component decreases to 0.07\% for the second outermost bin. Without XSG2, this bin shows no fraction of young stars, which indicates that this galaxy also has a larger contribution of young stars at larger radii. Finally, we also investigate the young stellar component from the individual galaxy spectrum of its central region. XSG2 hosts a f$\rm_{young}$=1.1\%. within its 0.8 kpc, which is larger than the fractions derived for our main sample (see Table \ref{tab:centralspectra}).

\begin{table}
 \centering
 \begin{tabular}{lccc}
 \hline	
Bin (kpc) & Age$\rm_{old}$ (Gyr) & [M/H]$\rm _{old}$ &f$\rm_{young}$  \\
  (1) & (2) & (3) & (4) \\
 \hline
0.0 -- 0.8  & 11.6$^{+0.5}_{-0.3}$ & 0.30$\pm0.02$ & 0.75$^{+0.09}_{-0.10}$\\
0.8 -- 1.2 & 11.9$^{+0.4}_{-0.3}$ & 0.20$\pm0.02$  & 0.27$\pm0.05$\\
1.2 -- 1.7 & 12.6$^{+0.5}_{-0.4}$ & 0.14$\pm0.03$ & 0.17$\pm0.03$ \\
1.7 -- 2.4 & 11.4$^{+0.6}_{-0.5}$ & 0.14$\pm0.03$  & 0.07$\pm0.03$ \\
2.4 -- 4.2 & 10.5$^{+2.0}_{-1.0}$ & 0.13$\pm0.05$ & 0 \\
 \hline
XSG2 (centre) & 11.75$^{+0.06}_{-0.03}$ & 0.22$\pm{0.01}$ & 1.10$\pm{0.02}$ \\	
 \hline
 \end{tabular}
\caption{Best-fit values of the stellar population parameters for each galactocentric distance bin when XSG2 is included in the stacked spectra. Last row shows the estimates for the XSG2 galaxy of the individual spectrum at the innermost region (<0.8 kpc). Ages and metallicities for the old stellar component are listed in the columns 2 and 3, respectively, and fraction of young stars in the column 4.
   }
 \label{tab:xsg2}
\end{table}


\bsp	
\label{lastpage}
\end{document}